\documentclass[11pt]{article}
\usepackage{amssymb}
\usepackage{latexsym}
\usepackage[mathscr]{eucal}
\usepackage{citesort}
\usepackage{url}
\usepackage{a4}
\usepackage{cite}
\newcommand{\hpsi}{\hat \psi}

\newcommand{\fcb}{\mathrm{fcb}}
\newcommand{\fcbe}{future \cbe}
\newcommand{\fcbes}{future \cbes}
\newcommand{\cbe}{conformal boundary extension}
\newcommand{\cbes}{conformal boundary extensions}
\newcommand{\spa}{{\mathrm{space}}}
\newcounter{mnotecount}[section]

\newcommand{\id}{{\mbox{\rm Id}}}
\newcommand{\mM}{{\mathring{M}}}

\newcommand{\mcN}{{\mycal N}}

\def \Reel{\mathbb{R}}
\def \T{\mathbb{T}}
\def \R {\Reel}

\def \Nat{\mathbb{N}}

\def \N {\Nat}

\newcommand{\be}{\begin{equation}}
\newcommand{\ee}{\end{equation}}
\newcommand{\bel}[1]{\begin{equation}\label{#1}}
\newcommand{\beal}[1]{\begin{eqnarray}\label{#1}}
\newcommand{\beadl}[1]{\begin{deqarr}\label{#1}}
\newcommand{\eeadl}[1]{\arrlabel{#1}\end{deqarr}}
\newcommand{\eeal}[1]{\label{#1}\end{eqnarray}}
\newcommand{\eead}[1]{\end{deqarr}}
\newcommand{\eea}{\end{eqnarray}}
\newcommand{\eeaa}{\end{eqnarray*}}

\newcommand{\eq}[1]{(\ref{#1})}

\DeclareFontFamily{OT1}{rsfs}{}
\DeclareFontShape{OT1}{rsfs}{m}{n}{ <-7> rsfs5 <7-10> rsfs7 <10->
rsfs10}{} \DeclareMathAlphabet{\mycal}{OT1}{rsfs}{m}{n}
\def\scri{{\mycal I}}%
\def\Scri{\scri}

\newcommand{\mcO}{{\mycal O}}

\newcommand{\mcU}{{\mycal U}}

\newcommand{\qed}{\hfill $\Box$\bigskip}

\newcommand{\proof}{\noindent {\sc Proof:\ }}

\newtheorem{defi}{\sc Definition\rm}[section]

\newtheorem{prop}[defi]{\sc Proposition\rm}

\newtheorem{Definition}[defi]{\sc Definition\rm}

\newtheorem{Theorem}[defi]{\sc Theorem\rm}

\newtheorem{Proposition}[defi]{\sc Proposition\rm}

\newtheorem{Lemma}[defi]{\sc Lemma\rm}

\newtheorem{Remark}[defi]{\sc Remark\rm}

\newcommand{\levoca}[1]{}

\begin{document}
\title{Conformal boundary extensions of Lorentzian manifolds}
\author{
Piotr T. Chru\'sciel\thanks{Email \protect\url{
Piotr.Chrusciel@lmpt.univ-tours.fr}, URL \protect\url{
www.phys.univ-tours.fr/}$\sim$\protect\url{piotr}}\\ F\'ed\'eration Denis Poisson, LMPT, Tours, and \\
Albert Einstein
Institute,  Golm\thanks{Visiting fellow.} %
}
\date{}

\maketitle

\begin{abstract}
We study the question of local and global uniqueness of
completions, based on null geodesics, of Lorentzian manifolds. We
show local uniqueness of such boundary extensions. We give a
necessary and sufficient condition for existence of unique maximal
completions. The condition is verified in several situations of
interest. This leads to existence and uniqueness of maximal
spacelike conformal boundaries, of maximal strongly causal
boundaries, as well as uniqueness of \cbes\ for asymptotically
simple space-times. Examples of applications include the
definition of mass, or the classification of inequivalent
extensions across a Cauchy horizon of the Taub space-time.
\end{abstract}

\section{Introduction}

 In
general relativity one often faces the need of extending the
space-times under consideration. For example, one might wish to
extend a globally hyperbolic space-time by adding a Cauchy
horizon~\cite{Misner}. Or one might wish to extend a domain of
outer communication, given in some coordinate system, by adding an
event horizon~\cite{ScottSzekeresII,Kruskal,Szekeres}. Finally,
when studying the asymptotic behavior of the fields, one might
wish to add to the space-time a conformal boundary at
infinity~\cite{penrose:scri}. In all those cases one makes a
``\cbe" of the
original Lorentzian manifold (in the first two cases the conformal
factor being one).

It is then natural to raise the question of uniqueness of the
extensions. It is the object of this paper to establish
some results concerning this problem. We start by proving
uniqueness of differentiable structure of the boundary extensions;
this is a purely local result. We then pass to an analysis of the
global aspects of the problem. We give a necessary and sufficient
condition for existence of unique maximal completions in terms of
families of null geodesics. We check that our condition is
verified in several situations of interest. This leads to a proof
of existence and uniqueness of maximal spacelike conformal
boundaries. We also obtain uniqueness of \cbes\ for asymptotically
simple space-times.

The reader is referred
to~\cite{GerochEspositoWitten,GKP,ChRendall,Schmidtcqg91,ChHerzlich,Schmidt74},
and also to~\cite{Harrisreview} and references therein, for
previous results related to the problems at hand.

This paper is the result of a collaboration with Robert Geroch, who
contributed several key ideas, and drafted parts of the text.

\section{Preliminaries}

 Let $M$ be a smooth $n$--manifold (without boundary), and
$g_{ab}$ a smooth, Lorentz-signature metric thereon. (Manifolds are
assumed to be paracompact and Hausdorff throughout.)

We will use the convention that a manifold with boundary $\tilde
M$ contains its boundary $\partial \tilde M$ as a point set
(recall that it is sometimes convenient, for PDE considerations,
not to do that). With this convention, spaces of functions such
as, \emph{e.g.},\/ $C^k(\tilde M)$ consist of functions which are
$k$ times continuously differentiable in the interior of $\tilde
M$, with the derivatives extending by continuity to continuous
functions on $ \tilde M$.

Let $\mcN_pM\subset T_pM$ denote the collection of null vectors at
$p$, and let $\mcN M \subset TM$ be the bundle of all null
vectors.

Denote by $B\subset TM$ the collection of (nonzero) null vectors
$\mu\in \mcN M$ such that $\exp ( \mu)$ exists. Then $B$ is a
smooth $(2n-1)$--manifold (arising as a smooth submanifold of the
tangent bundle).  Also, the exponential map $\exp$, when
restricted to $B$, is a smooth map from $B$ to $M$.

We restrict attention to $(M,g)$'s which are time oriented, and to
those null geodesics which are future directed.

We shall say that a Jacobi field is  \emph{null-connecting} if it
arises from a one-parameter differentiable family of \emph{null}
geodesics. (This should not be confused with the usual ``quotient
modulo the field of tangents" for Jacobi fields along null
geodesics; no such quotient will be taken here.)

 Consider, then, a differentiable
one-parameter family $\gamma(\lambda,s)$ of null geodesics, so
that for every $\lambda$ the map $\gamma(\lambda,\cdot)$ is an
affinely parameterised null geodesic, set $l=\gamma_*\partial_s$,
$X=\gamma_*\partial_\lambda.$ It is well known (and in any case
easily follows from the vanishing of the torsion) that
$\nabla_Xl=\nabla_l X$. Since $g(l,l)=0$ for all geodesics in the
family under consideration, one finds
$$0=X(g(l,l))=2g(l,\nabla_Xl)=2g(l,\nabla_lX)=2l(g(l,X))\;.$$
It follows that for all null-connecting Jacobi fields $X$ along a
null geodesic $\gamma(s)$ we have
\bel{nceq} \forall \ s_1, s_2 \quad g(l(s_1),X(s_1))=g(l(s_2),X(s_2))\;.\ee

We shall say that $y$ is \emph{null-conjugate} to $x$ if
$y=\exp(s_0\mu)$ with $s_0>0$ for some null vector $\mu\in T_xM$
and if there exists a null-connecting Jacobi field along  the
 geodesic segment $[0,s_0]\ni s\to \exp(s\mu)$ which \emph{vanishes precisely at
$x$ and at $y$}. Note that a null-conjugate point is necessarily a
conjugate point in the usual sense, but the inverse implication is
false in general.

 From \eq{nceq} we immediately obtain:

\begin{Proposition}
\label{Cl5n} Let $X$ be a null-connecting Jacobi field on
 a geodesic segment $\gamma$ with $X(0)$ --- timelike. Then $X$ has no zeros on $\gamma$.
\end{Proposition}

\proof Since $ X(0)$ is timelike we have $g(\dot \gamma
(0),X(0))\ne 0$, and the result follows from \eq{nceq}. \qed

Next, let $\kappa$ be a smooth timelike curve in $M$, note that
$\dot \kappa$ has no zeros. Denote by $S_\kappa\subset B$ the
collection of those vectors $\mu\in \mcN M$ such that i)
$\pi(\mu)\in\kappa$ (by an abuse of notation, we will use the same
symbol $\kappa$ for a curve and for  its image), where $\pi:\mcN
M\to M$ is the projection map, and ii) the null geodesic segment
\bel{Gmps} t\to \Gamma_{\mu}(t):= \exp(t\mu)\;,\quad t\in
[0,1]\;,\ee generated by $\mu$ has no null-conjugate points. Thus,
$S_\kappa$ is an open subset of the $n$-submanifold
$\pi^{-1}(\kappa)\subset B$ of $B$, hence a smooth $n$-submanifold
of $B$.

Any (non-trivial) flat Riemannian cone (without its tip)
multiplied by $(n-2)$--dimensional Minkowski space-time provides
an example of space-time in which $\exp|_{S_\kappa}$ might fail to
be injective for some timelike curves $\kappa$. We note that there
are no null-conjugate points in this space-time.

\begin{Proposition}
\label{Pkern}The map $$S_\kappa \stackrel{\exp}{\rightarrow}
\exp(S_\kappa) \subset M$$ is a submersion.

\end{Proposition}

\proof The usual analysis of Jacobi fields shows that $\mbox{\rm
Ker} \left((\exp|_{S_\kappa})_* \right)\ne \{0\}$ if and only if
there exist null-connecting Jacobi fields along $\Gamma_{\mu}$
which vanish at $\exp(\mu)$ and which are initially tangent to
$\kappa$, possibly vanishing there. If $X(0)$ vanishes, then the
fact that  $X$ has no other zeros follows from the definition of
$S_\kappa$. If $X(0)\ne 0$, then $X(0)$ is timelike since it is
tangent to $\kappa$, and the fact that $X$ is nowhere vanishing
follows from Proposition~\ref{Cl5n}. \qed

Let $S'_\kappa$ denote any open subset of  $S_\kappa$ with the
property that $\exp|_{S'_\kappa}$ is injective.  It then follows
from Proposition~\ref{Pkern} that $\exp|_{S'_\kappa}$ is a
diffeomorphism into its image.

 Choose any such $S'_\kappa$ and denote by $f$
the function on the $n$--manifold $S'_\kappa$ given by
``$\kappa$-parameter value": if $s$ is a parameter along $\kappa$,
then \bel{timfdef} f(\kappa(s),\mu):=s\;;\ee so $f$ is smooth and
has nowhere vanishing gradient. Its composition with
$(\exp|_{S'_\kappa}) ^{-1}$ (resulting in a function on
$S'_\kappa$), then, is also smooth on its domain of definition and
has nowhere vanishing gradient there. In fact, we have the
following:

\begin{Proposition}
\label{Pgrad} The gradient of
\bel{gdef}u_\kappa:=f\circ\,\exp^{-1}\ee at\/ $\exp( \mu) \in \exp
(S'_\kappa)\subset M$ is proportional to the tangent, at this
point, to the geodesic generated by $\mu $, with non-zero
proportionality factor.
\end{Proposition}
 \begin{Remark}
\label{Rdifferentiability} Some comments about differentiability
thresholds are in order: our analysis requires local uniqueness of
geodesics, as well as controlled behavior of Jacobi fields. Those
properties will hold if the metric is $C^{k,1}$, with $k\ge 1$.
Such differentiability of the metric leads naturally to
$C^{k+1,1}$ differentiability of the manifold. The exponential map
is then a $C^{k-1,1}$ map (both as a function of $\mu\in
T_{\pi(\mu)}M$ and of $\pi(\mu)$), while Jacobi fields are of
$C^{k-2,1}$ differentiability class (since the Jacobi equation
involves the Riemann tensor) as functions of initial data at $p\in
M $ and of $p$. Because Jacobi fields govern the differentiability
of families of geodesics, the field of tangents to the level sets
of $u$ will be $C^{k-2,1}$ differentiability class. This results
in $C^{k-1,1}$ differentiability of $u_\kappa$ when $\kappa$ is
$C^{k-1,1}$ or better. This is a loss of two derivatives as
compared to the expected differentiability class of $M$.

\end{Remark}

\begin{Remark}
\label{Rphg} We also note that, for solutions of field equations,
a possible  hypothesis on the metric near a conformal boundary at
infinity \emph{\`a la Penrose} is that of \emph{polyhomogeneity},
see e.g.~\cite{ValienteKroon:2003ux,ChMS}. If $\kappa$ is smooth,
then $u_k$ will be polyhomogeneous and $C^{k+1}$ at the conformal
boundary whenever the metric is polyhomogeneous and $C^k$ there,
$k\ge 1$. (The increase of differentiability, as compared to
Remark~\ref{Rdifferentiability}, arises from the fact that
integrals, along curves which are transverse to the boundary, of
polyhomogeneous functions increase differentiability by one; this
gives $C^k$ and polyhomogeneous Jacobi fields, hence a $C^k$ field
of null normals at the conformal boundary, hence a $C^{k+1}$ and
polyhomogeneous function $u_\kappa$ by the analysis of
~\cite[Appendix~B]{ChMS}.)
\end{Remark}

\proof The function $f\circ\,\exp^{-1}$ is constant along the
future light cones issued from $x(s)\in \kappa$, hence its
gradient is proportional to the normal to those light cones. Since
the subsets of those light cones which lie in the image by $\exp$
of $S'_\kappa$ form smooth null hypersurfaces, their normals are
proportional to the tangents of their geodesic generators. The
fact that the proportionality factor is non-zero follows from the
fact that $\exp|_{S'_\kappa}$ is a diffeomorphism into its image
and that $f$ has non-vanishing gradient on ${S'_\kappa}$. \qed

The functions $u_\kappa$ as in \eq{gdef} provide the main tool for
our local analysis below.

Next, let $L$ be a smooth $(n-1)$--submanifold of $M$.
 We will actually be mainly interested in situations when
$L$ is a boundary of $M$; in order to make this compatible with
our hypotheses above,  we then extend $M$ to a new manifold, still
denoted by $M$. Thus, without loss of generality, we can assume
that $L$ is an interior submanifold of $M$; the discussion above
is independent of the extensions done. Denote by $B_L\subset B$
those points $\mu $ such that i) $\exp \mu \in L$, and ii) the
null geodesic generated by $\mu $ meets $L$ transversally (with
\emph{all} intersections transverse, whenever there is more than
one). Then $B_L$ is a smooth $(2n-2)$--(sub)manifold (of $B$). Fix
$\mu \in B_L$. Denote by $\Gamma$ the tangent, at $\mu $, to the
curve in $B$ given by $\exp( a\mu)$, with $a$ in the domain of
definition of $\exp(\cdot)$. Then, by transversality, $\Gamma$ is
not tangent to the submanifold $B_L$. That is, this curve meets
$B_L$, at $a = 1$, transversally. So, since $B_L$ has codimension
one in $B$, all nearby such curves (\emph{i.e.},\/ $\exp(a\mu')$,
for $\mu'$ near $\mu$) also meet $B_L$, and also transversally. In
other words, all null geodesics generated by elements near $\mu $
meet the original submanifold $L$, also transversally.

\section{Local uniqueness of \cbes\  }

The following result establishes uniqueness of the differentiable
structure of \cbes\   under suitable hypotheses:

\begin{Theorem} \label{T1} Let $M$ be a
smooth $n$--manifold, $n\ge 3$, and $g_{ab}$ a smooth
Lorentz-signature metric thereon. Let $(\hat{M}, \hat{g}_{ab})$ be
a smooth $n$--manifold with boundary with smooth Lorentz-signature
metric, and let $M\stackrel{\psi}{\rightarrow}\hat{M}$ be a smooth
diffeomorphism into, such that i) $\psi[M] = \hat{M}
\setminus\partial \hat{M}$, and ii) the $\psi$-image of $g_{ab}$
is conformal to $\hat{g}_{ab}$. Similarly for $\hat{M}'$,
$\hat{g}'_{ab}$, and $\psi'$.  Let
$I\stackrel{\gamma}{\rightarrow} M$, where $I$ is an open interval
in $\R$, be a directed null geodesic in $(M, g_{ab})$ such that
$\hat{\gamma} = \psi\circ \gamma$ has future endpoint $p\in
\hat{M}$ and $\hat{\gamma}' = \psi'\circ \gamma$ has future
endpoint $p'\in \hat{M}'$.  We also assume  that $\hat{\gamma}$
meets $\partial \hat{M}$, and $\hat{\gamma}'$ meets
$\partial[\hat{M}']$, transversally. Then there exist open
neighborhoods $U$ of $p$ in $\hat{M}$ and $U'$ of $p'$ in
$\hat{M}'$, together with a smooth diffeomorphism $U
\stackrel{\phi}{\rightarrow} U'$, such that I) $\phi(p) = p'$, and
II) restricted to $U\cap \psi[M]$, $\psi'\circ\psi^{-1} = \phi$.
\end{Theorem}

\begin{Remark} The result is wrong in
space-time dimension two. In order to see that, let $M$ be the
$2$-manifold with chart $(u,v)$ where $u < 0$, $v \in \R$. Let $g$
be the metric $dudv$ on $M$. Let $\gamma$ be the null geodesic
given by $u = \lambda, v = 0$, where $\lambda$ (the parameter)
goes from $-\infty$ to $0$. We introduce two conformal boundary
completions in which $\gamma$ acquires an endpoint.

We let, first, $M'$ be the obvious completion: thus $M'$ is the
manifold with boundary with chart $(a,b)$, where $a \le 0$, $b \in
\R$; with metric $dadb$. Let $\psi'$ be the mapping from $M$ to
$M'$ that sends $(u,v)$ to the point $a = u$, $b = v$. This is a
\cbe\ (with conformal factor one) terminating $\gamma$; by this we
mean that $\gamma$ has an end point at the boundary.

We let, next, $M''$ be the manifold with boundary with chart
$(p,q)$, where $p \le 0$, $q \in \R$; with metric $dpdq$.  Let
$\psi''$ be the mapping from $M$ to $M''$ that sends $(u, v)$ to
the point $p = - \sqrt{-u}$, $q = v$. This is again a \cbe\ (with
non-trivial conformal factor), terminating $\gamma$.

Clearly the map $\psi''\circ (\psi')^{-1}$ does not satisfy the
conclusions of Theorem~\ref{T1}.
\end{Remark}

\begin{Remark}
The main situations of interest in general relativity are  with a
conformal factor identically one, or with a conformal factor
relating $g$ and $\hat g$ that degenerates at $\partial \hat M$ to
first order; similarly for $\hat g'$. We emphasise that no such
restrictions are made here. (On the other hand, we expect that in
dimension two  the proof of Theorem~\ref{T1} can be adapted with
such supplementary assumptions.)
\end{Remark}

\begin{Remark}
\label{Rdiff} If $C^{k+1,1}$ differentiability of  the manifolds
and $C^{k,1}$ differentiability of  the metrics involved is
assumed, $k \ge 1$, then the result remains true with $\phi$ of
$C^{k-1,1}$ differentiability class, see
Remark~\ref{Rdifferentiability}. On the other hand, if the metric
is assumed to be polyhomogeneous and $C^k$, $k\ge 1$, near the
conformal boundary, then $\phi$ will be polyhomogeneous and
$C^{k+1}$, see Remark~\ref{Rphg}.
\end{Remark}

\proof Let $q$ be a point of $\hat{\gamma}$, $\lambda$ the
direction of the curve $\hat{\gamma}$ at $q$, $A$ an open
neighborhood of $\lambda$ in the $(2n-2)$--manifold of null
directions at points of $\hat{M} \setminus\partial \hat{M}$, and
$U$ the (open) subset of $\hat{M}$ consisting of the union of null
rays generated by all points of $A$.  Then, we claim,

\begin{prop}\label{Pclose}
For all $q$ sufficiently close to $p$, and all $A$ sufficiently
small, there exists at most one $\hat g$--geodesic between any two
points in $U$, and:
\begin{enumerate} \item The spacetime $U$ is strongly
causal, \item every null ray generated by a point of $A$ has
future endpoint on $\partial \hat{M}$ and meets $\partial \hat{M}$
transversally there, and
\item no null geodesic segment contained in $U$ has a pair of
conjugate points in $(U,\hat g|_{U})$.
\end{enumerate}
\end{prop}

\proof Let $O$ be a geodesically convex neighborhood  of $p$.

1. We take a strongly causal neighborhood of the point $p$ which
is included in $O$. Given any $q$ such that the segment $[q,p]$ of
$\hat{\gamma}$ lies entirely within that neighborhood and any $A$
sufficiently small we will have that $U$ lies entirely within that
neighborhood.

2. This follows  from transversality of $\hat \gamma$ at $\hat p$
and from continuous dependence of solutions of ODE's upon initial
values.

3.  Let $\hat h$ be any auxiliary Riemannian metric on $\hat M$;
the result follows from the fact that the uniform bound on the
Riemann tensor of $\hat g$ on a compact set allows one to control
from below the $\hat h$-distance to a conjugate point. \qed

Let $\hat q= \psi(q)$ be close enough to $p$ so that the
conclusions of Proposition~\ref{Pclose} hold for some open
neighborhood  $\hat A=\psi_*(A)$ of $\hat \lambda=\psi_*
\lambda\in T_{\hat q} \hat M$, as determined by the boundary
$\partial \hat M$ in the space-time $(\hat M,\hat g)$, and that
the same conclusions apply to some open neighborhood $\hat A'=
\psi'_*(A')$ as determined by the boundary $\partial \hat M'$ in
the space-time $(\hat M',\hat g')$. Replacing $A$ by $A\cap A'$
one then finds that the conclusions of Proposition~\ref{Pclose}
simultaneously hold for $\psi_*(A)$ and for $\psi'_*(A)$, with
sets $\hat U\subset \hat M$ and $\hat U'\subset\hat M'$.

 We let $\pi$ denote
the projection from $TM$ to $M$.

Replacing $\hat M$ by $U$, we can without loss of generality
assume that $\hat M$ is strongly causal. This implies that, close
enough to $p \in \hat M$, the boundary $\dot J^-(p)$ of the causal
past of $p$, after removal of  $p$, is a smooth submanifold of
$\hat M$.

We wish to construct convenient coordinate systems near the
boundary, the following assertion will be useful:

\begin{Proposition}\label{Pobv} Let $n\ge 3$ and let $\mcU$ be a neighborhood of $p\in \partial \hat M$ such that
$(\dot J^-(p)\cap \mcU)\setminus \{p\}$ is a smooth submanifold.
Denote by $\tilde K$ the collection of all null vectors $ l$, with
$\pi(l)\in \psi^{-1}(\dot J^-(p)\cap \mcU)\subset M$,  such that
the corresponding null geodesic in $\psi(M)=\hat M
\setminus\partial \hat M$ ends at $p\in\partial \hat M$, and there
its tangent is transverse to the boundary of $\hat M$. Let $K$
denote the connected component of $\tilde K$ containing  all the
null vectors $\dot \gamma$ (where $\gamma$ is as in the statement
of Theorem~\ref{T1}), and let $W$ be the subset of
$$\underbrace{K \times \ldots\times K}_{\mbox{\scriptsize $n$
factors}} $$ consisting of $n$-tuples such that the null-tangents
to the corresponding null geodesics are linearly independent at
$p$. 
Then $W$ is open and dense in $K \times \ldots \times K $.
\end{Proposition}

\proof Let $\alpha:K\to T_p\hat M$ denote the map which to a
vector $l\in K\subset T_qM$ assigns the tangent at $p$ to the
corresponding affinely parameterised null geodesic passing through
$q$, with tangent $l$ there. Then $\alpha$ is  continuous  on $K$.
Let $(\alpha(l_1) \ldots \alpha(l_n))$ denote the matrix obtained
by juxtaposition of the vectors $\alpha(l_i)$, then $W=\{\det
(\alpha(l_1) \ldots \alpha(l_n))\}\ne 0$, hence $W$ is open.

Consider any point $(l_1,\ldots, l_n)$ in $K\setminus W$, and let
$k<n$ be the rank of $(\alpha(l_1) \ldots \alpha(l_n))$. By
reordering we can without loss of generality assume that
$\{\alpha(l_1), \ldots, \alpha(l_k)\}$ are linearly independent.
Suppose that there exists an open set $\mcO\subset M$ around
$q_{k+1}=\pi(l_{k+1})$ such that $\alpha(x)\in \mbox{\rm
Vect}\{\alpha(l_1), \ldots, \alpha(l_k)\}$ for all $l\in K\cap
\pi^{-1}(\mcO)$, then $\pi (K)\cap \mcO$ is a subset of the smooth
$(k-1)$--dimensional manifold
$$\psi^{-1}\left(\{\exp(m),\; m \in \mbox{\rm Vect}\{\alpha(l_1), \ldots,
\alpha(l_k)\},\; g(m,m)=0\}\cap \mcU\right)\;.$$ However, $\pi
(K)\cap \mcO=\dot J^-(p)\cap \mcO$; since $\dot J^-(p)$ is an
achronal boundary, the set $\dot J^-(p)\cap \mcO$ is  an
$(n-1)$--dimensional topological submanifold, which gives a
contradiction. It follows that there exists an
$\epsilon$-perturbation of $l_{k+1}$ such that the rank of
$(\alpha(l_1) \ldots \alpha(l_n))$ at the perturbed points $l_i$
is $k+1$. In a finite number of steps one obtains a perturbation,
as small as desired, such that the rank of $(\alpha(l_1) \ldots
\alpha(l_n))$ at the perturbed points $l_i$ is $n$. \qed

Let $\mcU'$ be a set in $\hat M'$ as in Proposition~\ref{Pobv},
passing to subsets we can assume that
$\psi'{}^{-1}(\mcU')=\psi^{-1}(\mcU)$.
 Let $K'$ be the analogous collection as determined by $p'$
and $\hat M'$; it is not completely clear whether or not $K=K'$,
however, $K\cap K'$ forms a neighborhood, both within $K$ and
$K'$, of all the points $\dot \gamma(s)$. Let $W'$ be the
corresponding $W$ set, then $W\cap W'$ is still open and dense in
$(K\cap K') \times \ldots \times (K\cap K') $, further the
null-tangents to the corresponding null geodesics are now linearly
independent both at $p\in\partial \hat M$ and at $p'\in
\partial \hat M'$.

Let $$\left(\ell(1),\ldots, \ell(n)\right)\in W\cap W'\cap
(A\times \ldots \times A)\;$$ this last set  is not  empty in view
of what has been said. Choose $n$ smooth, timelike curves
$\kappa(a)$, $a=1,\ldots,n$, in $M$ such that $\kappa(a)$ passes
through $q(a)$. The construction before Proposition~\ref{Pgrad}
applies to $(\hat U,\hat g|_{\hat U})$ and leads to $n$ smooth
functions $u _{\psi\circ\kappa(a)}$, associated with the timelike
curves $\psi\circ\kappa(a)$, defined throughout $\hat U$. One can
likewise carry that construction on $(\hat U',\hat g'|_{\hat
U'})$, using the curves $\psi'\circ\kappa(a)$, obtaining $n$
functions $u _{\psi'\circ\kappa(a)}$. Clearly \bel{good} u
_{\psi\circ\kappa(a)}\circ \psi = u _{\psi'\circ\kappa(a)}\circ
\psi'\;.\ee By construction the gradients of the $u
_{\psi\circ\kappa(a)}$'s are linearly independent at $p$,
similarly  the gradients of the $u _{\psi'\circ\kappa(a)}$'s are
linearly independent at $p'$. Passing to subsets of $\hat U$ and
$\hat U'$ if necessary, by continuity those gradients will be
linearly independent throughout $\hat U$ and $\hat U'$. It follows
that $\{u _{\psi\circ\kappa(a)}\}_{a=1,\ldots,n}$ forms an
$\hat{M}$-admissible coordinate system in a neighborhood of $p$,
similarly for $\{u _{\psi'\circ\kappa(a)}\}_{a=1,\ldots,n}$. The
map $\phi$ is defined to be the identity in the coordinate systems
just constructed. It should be clear that I) holds, while II) is
equivalent to \eq{good}. \qed

\section{Maximal extensions}\label{Sglobal}

 Suppose that a
manifold without boundary $(M,g)$ has been conformally extended to
a manifold with boundary $(\hat M,\hat g)$. One can then always
produce a smaller extension by removing points from $\partial \hat
M$. This is a rather trivial operation, which can be reverted by
adding back the points that have been discarded. One is thus led
to the notion of \emph{maximal extension}: this is an extension
which cannot be ``made larger by adding  points". Then the
question arises, how many maximal extensions exist. This is the
issue that we want to address in this section. We start by
developing some terminology, and tools, which will be used to
handle the problems at hand.

Fix, once and for all, a smooth $n$-dimensional manifold $M$
(without boundary), and a smooth, Lorentz-signature metric, $g$,
thereon. By a {\em \cbe} of a pseudo-Riemannian manifold $(M, g)$,
we mean a triple, $\alpha' = (M', g', \psi')$, where $M'$ is a
smooth manifold (possibly) with boundary, $g'$ is a smooth,
Lorentz-signature metric on $M'$, and $\psi'$ is a smooth mapping,
$M\stackrel{\psi'}{\rightarrow}M'$, such that $\psi'$ i) is a
diffeomorphism into, ii) satisfies $\psi'[M] = M' - \partial M'$,
with iii)  $g$ being conformal to  $(\psi')^* g'$.

We note that the question of the \emph{behavior of the relative
conformal factor when the boundary is approached is completely
irrelevant} from our point of view, so our analysis covers the
usual Penrose-type situations where the inverse of the conformal
factor vanishes throughout the boundary (we will talk about
\emph{conformal completions at infinity} in this case), but also
situations where the conformal factor equals one, but also
situations where the conformal factor tends to any values in
$\R^+\cup\{\infty\}$ as one approaches the boundary, or perhaps
does not even have  limits at the boundary points.

When we write $\alpha''$, the associated objects  will  be
implicitly  understood as $(M'',g'',\psi'')$, similarly for
$\alpha_a$, $\hat \alpha$, \emph{etc.}

A \cbe\ $\alpha'$ will be called \emph{future} if every point of
$\partial M'$ is the future end point of some null geodesic;
similarly one can talk about \emph{past} \cbes. It is easily seen
that every point in a future extension terminates \emph{some} null
geodesic transversally. Note that a future \cbe\ can
simultaneously be a past one.

From now on \emph{we restrict attention to future extensions}. The
 analysis below applies to past extensions, or to extensions, after obvious modifications,
 using past inextendible, or inextendible, null geodesics. Note,
 however, that there are situations where the constructions
 below using simultaneously future and past extensions yield
 an empty maximal boundary, while non-trivial future and past
 boundaries exist.

Given a manifold with boundary $M$, we use the symbol $\mathring
M$ to denote its interior, $\mathring M:= M \setminus
\partial M$.

 Given any two \cbes\
$\alpha'$ 
and $\alpha''$ 
we write $\alpha' \leq \alpha''$ provided there exists a smooth
conformal diffeomorphism into $M'\stackrel{\phi}{\rightarrow} M''$
satisfying $\phi \circ \psi' = \psi''$. We write $\alpha'
\leq_\phi \alpha''$ when the map $\phi$ has been chosen.

We write $\alpha' \sim \alpha''$ if there exists a $\phi$ as above
which is a diffeomorphism \emph{of manifolds with boundary}. This
is obviously an equivalence relation.

The reader should note that two equivalent extensions are
conformal to each other, but sometimes extensions which are
conformal to each other might fail to be equivalent. The following
simple example illustrates this: Let $M$ have coordinates $(u, v)$
each with range $(-1,1)$, with metric $du \;dv$.  Let $M'$ have
global coordinates $$(u', v')\in(-1, 1]\times
(-1,1]\setminus\{(1,1)\}\;,$$ with metric $du'\; dv'$.  Let $
\psi'$ send $(u,v)$ in $M$ to $u' = u, v' = v$ in $M'$. Let $M''$
have global coordinates $$(u', v')\in[-1, 1)\times
[-1,1)\setminus\{(-1,-1)\}\;,$$ with metric  $du''\; dv''$. Let,
finally, $\psi''$ send $(u, v)$ in $M$ to $u'' = u, v'' = v $ in
$M''$. So, we now have a space-time, and two conformal boundary
extensions $\alpha'$ and $\alpha''$. Clearly $(M,g')$ is isometric
with $(M'',g'')$, the relevant isometry being the map which sends
$(u',v')$ to $(u'',v'')=(-u',-v')$.  However,
$\alpha'\not\sim\alpha''$. Indeed, suppose that there exists
$\psi$ such that $\psi\circ \psi'=\psi''$, then for $(u,v)\in
\mathring M'$ we have
$$\psi(u,v)=\psi\circ \psi'(u,v)=\psi''(u,v)=(u,v)\;,$$
so that $\psi(u,v)=(u,v)$, which does not extend to a map from
$\partial M'$ to $M''$.

A less trivial example of the above behavior is given by the two
standard extensions of the Taub region of Taub-NUT space-time,
which again are isometric but nonequivalent.

 In our new terminology,
Theorem~\ref{T1} is equivalent to the following statement:

\begin{Theorem}
Let $(M, g)$ be a spacetime, $\gamma$ a future-inextendible null
geodesic therein, and $\alpha$ and $\alpha'$ conformal extensions
in each of which $\gamma$ acquires a transversal future endpoint.
Then there exists a conformal extension, $\alpha''$, which is
smaller than both $\alpha$ and $\alpha'$, and in which $\gamma$
also acquires a transversal future endpoint.
\end{Theorem}

It follows immediately from the definitions above that if $\alpha'
\leq_{\phi_1} \alpha''$ and $\alpha'' \leq_{\phi_2} \alpha'$, then
$\phi_1$ and $\phi_2$ are inverses of each other \emph{on the
interiors $\mathring M'$ and  $\mathring M''$}:
\bel{intid}\phi_1\circ \phi_2\Big|_ {\mM'}= \id_{\mM'}\;,\quad
\phi_2\circ \phi_1\Big|_ {\mM''}= \id_{\mM''}\;.\ee Because both
$\phi_1$ and $\phi_2$ smoothly extend by continuity to the
boundaries of their domains of definition by hypothesis, we claim
that
$$\phi_1\circ \phi_2= \id_{M'}\;,\quad \phi_2\circ \phi_1=
\id_{M''}\;.$$ Indeed, consider a sequence $p_n\in \mM'$ such that
$p_n\to p\in \partial M'$, then $\phi_2(        p_n)$ converges to
$\phi_2(p)$, so that the left-hand-side of \eq{intid} converges to
$\phi_1(\phi_2(p))$, while the right-hand-side converges to $p$.
We have therefore obtained:
\begin{Proposition}\label{Ppartialor}
The relation $\leq$  is a partial ordering on the set of
$\sim$--equivalence classes (we will still use the same symbol
$\leq$ for that new relation).
\end{Proposition}

 We want to construct maximal future
 \cbes. However, the Taub-NUT example shows that completions
 obtained by ``attaching everything that can be attached"
 might lead to non-Hausdorff behavior. In order to gain insight in the
phenomena that occur, some further terminology will be needed:

\begin{Definition} \label{Dugly} Let
$\gamma_1$ and $\gamma_2$ be null geodesics which are maximally
extended to the future.
\begin{enumerate}\item  For $x\in M$ and $l$ a null vector at $x$, by a
{\em thickening} of the corresponding null geodesic we mean the
union of all null geodesics arising from initial conditions in a
($ TM$--)neighborhood  of $l$.
\item \label{P2}We shall say that the pair
$(\gamma_1,\gamma_2)$ is \emph{simple} if  i) each null geodesic
enters and remains in every thickening of the other; or ii) after
moving far enough along each null geodesic one can find
thickenings $\mcO_a$ of each, $a=1,2$, that have empty
intersection, $\mcO_1\cap \mcO_2=\emptyset$.

\item We shall say that  $\gamma_1$ and $\gamma_2$ are
\emph{intertwined} if $(\gamma_1,\gamma_2)$ is not simple.


\item A collection $\Omega$ of null future inextendible geodesics
will be called \emph{simple} if  $\Omega$ is open and if all pairs
$(\gamma_1,\gamma_2)\in \Omega\times \Omega$ are simple.

\item A simple collection $\Omega$ of null future inextendible geodesics
will be called \emph{maximal} if for any open set $\Omega_1$ of
future inextendible null geodesics the collection $\Omega\cup
\Omega_1$ is not simple.
\end{enumerate}
\end{Definition}

An equivalent way of saying that two geodesics are intertwined is
the following: For any sufficiently small thickening of either,
the other null geodesic intersects that thickening but does not
remain in it.

The reader will easily check that any collection of future
inextendible null geodesics in Minkowski space-time is simple.
(This also follows from Theorems~\ref{Tnewmain2} and \ref{Tugly2}
below.) On the other-hand, the collection of all future
inextendible null geodesics in Taub-NUT space-time is not, compare
Section~\ref{SCCh}.

The set of simple collections of null geodesics is directed by
inclusion, and it is a simple consequence of the Kuratowski-Zorn
lemma that for every simple $\Omega$ there exists a simple maximal
$\hat\Omega$ such that $\Omega\subset \hat \Omega$. Examples, e.g.
in polarised Gowdy space-times with Cauchy horizons
(see~\cite{ChImaxTaubNUT}), show that there might be more than one
such $\hat\Omega$.

Conformal boundaries provide simple collections of null geodesics:

\begin{Proposition}
\label{POmst} Let $\hat \alpha$ be a \cbe\   of $(M,g)$, and let
$\Omega(\hat \alpha)$ denote the collection of maximally extended
null geodesics in $M$ which acquire a transverse future end point
on the conformal boundary $\partial \hat M$. Then ${\Omega(\hat
\alpha)}$ is simple.
\end{Proposition}

{\noindent\sc Proof:}  Openness of $\Omega(\hat \alpha)$ is
obvious. Let $\gamma_a$, $a=1,2$, be two null geodesics in
$\Omega(\hat \alpha)$, and let $\hat \gamma_a$ be their future
inexentendible extensions in $\hat M$. Then  the $\hat \gamma_a$'s
possess end points $p_a$ on the boundary $\partial \hat M$ of
$\hat M$, with null tangents
 $m_a\in T_{p_a} M$ there with respect to some convenient
 parameterisation; for instance, one can use the distance with respect
 to some chosen smooth-up-to boundary Riemannian  metric on $\hat M$. If
$p_1=p_2$ we are in case $i)$ of Definition~\ref{Dugly}.
Otherwise, we can find small disjoint neighborhoods $\mcU_a\subset
T\hat M$ of $(p_a,m_a)$ so that $\exp(\mcU_1)\cap
\exp(\mcU_2)=\emptyset$. (Here we view $\exp$ as a map from $T\hat
M$ to $\hat M$.) Then any thickening of $\gamma_a$ contained
within $\exp(\mcU_a)\cap M $ will satisfy the condition of case
$ii)$  of Definition~\ref{Dugly}. \qed

We are ready now to address the question, when does a family of
\cbes\   arise from a single larger one:

\begin{Theorem}\label{Tnewmain} Let $(M, g)$ be a space-time (without boundary), and
  consider any non-empty collection $\Xi$ of \fcbes.  There exists
  an
  extension $\alpha_\Xi$ that is larger than or equal every one of these if and only if
   no two of  extensions in $\Xi$
   terminate
  intertwined null geodesics.
\end{Theorem}

\begin{Remark}\label{Rnewmain}
The reader will notice that $M_\Xi$ constructed in the proof below
is the minimal one which is larger than all the extensions in
$\Xi$, hence unique.
\end{Remark}

\proof  The necessity of the condition follows from
Proposition~\ref{POmst}, it remains to show sufficiency.

For $\alpha' = (M', g', \psi')$ any \cbe\  in $\Xi$, denote by
$B_{\alpha'}$ the collection of all pairs, $(x, l)$, where $x \in
M$, and $l$ is a null vector in $M$ at $x$, such that:  The
affinely parameterised null geodesic in $M'$, with initial
condition the $\psi'$-image of $(x, l)$, meets the boundary of
$M'$, transversally. Then $B_{\alpha'}$ is open in the null bundle
of $M$.

Denote by $B_\Xi$ the union of the $B_{\alpha'}$, over all \cbes\
$\alpha'\in \Xi$. Call two points of $B_\Xi$ equivalent if there
exists a \cbe\   in which the two corresponding null geodesics (in
the \cbe\ manifold) meet at the same
 boundary point (of that \cbe\
manifold). This is an equivalence relation, the only non-trivial
property to check is transitivity: suppose, thus, that there
exists a completion $\alpha' = (M', g', \psi')$ in which the
maximal extensions of the null geodesics $\psi'(\gamma_1)$ and
$\psi'(\gamma_2)$ meet the boundary $\partial M'$ transversally at
$p'$, and  that there exists a completion $\alpha'' = (M'', g'',
\psi'')$ in which the maximal extensions of the null geodesics
$\psi''(\gamma_2)$ and $\psi''(\gamma_3)$ meet the boundary
$\partial M'$ transversally at $p''$. By Theorem~\ref{T1} the map
$\psi''\circ (\psi')^{-1}$ extends smoothly to $\partial M''$, so
that the maximal extension of the geodesic
$\psi''(\gamma_1)=\psi''\circ (\psi')^{-1}(\psi'(\gamma_1))$ meets
$\partial M''$ transversally at $p''$, and transitivity follows.

Set $M_\Xi$ the disjoint union of $M$ and the set of equivalence
classes. Now, given any \cbe\ in $\Xi$, $\alpha' = (M', g',
\psi')$, we consider the mapping
$M'\stackrel{\zeta'}{\rightarrow}M_\Xi$, given as follows. For $x$
any point of $M$, this map $\zeta'$ sends $\psi'(x)\in M'$ to the
point $x\in M$ on the left ($M$) side of the disjoint union that
is $M_\Xi$.  For $p \in M'$ on the boundary, $\zeta'$ sends $p$ to
the equivalence class of elements of $B_\Xi$ consisting of those
that generate the geodesics (in $M'$) that meet at the boundary
point $p$ of $M'$.  The images of these maps $\zeta'$ (for the
various \cbes) cover $M_\Xi$ by construction, as any boundary
point of a \cbe\ is the end point of some null transverse
geodesic. To see that the $\zeta'$'s are injective, let
$\zeta'(p_1)=\zeta'(p_2)$, thus there exists a \cbe\   $\alpha''$
and  null geodesics $\gamma_1$ and $\gamma_2$ which acquire ends
points $p_1$ and $p_2$ on $\partial M'$, and acquire the same end
point $p''$ on $\partial M''$. This last property implies that we
are in case $i)$ of point~\ref{P2} of Definition~\ref{Dugly}, and
it easily follows that $p_1=p_2$.

We now introduce charts on the set $M_\Xi$ by taking charts on
\cbes, and sending them to $M_\Xi$ via the map $\zeta'$. These
cover $M_\Xi$; any two of them arising from a single completion
are obviously compatible, while  any two arising from two
different completions can be seen to be compatible by
Theorem~\ref{T1}: Indeed, consider two coordinate charts, one
around a boundary point $p'$ in a completion $\alpha'$, and a
second one around a corresponding boundary point $p''$ in a
completion $\alpha''$. In order to apply Theorem~\ref{T1} we need
to find a null geodesic which will be transverse to the conformal
boundary in both
completions. Let $K'$ be 
the set  $K$ associated with $p'$, as defined in
Proposition~\ref{Pobv}. Then $K'$ is open and dense in the set of
all $(q,\ell)$  in $TM$ such that $\ell$ is null, and such that
the associated null geodesic ends at $p'$.  Let $K''$ be the set
$K$ of Proposition~\ref{Pobv} associated to $p''$, this is again
an open and dense subset in the set of all $(q,\ell)$  in $TM$
such that $\ell$ is null, and such that the associated null
geodesic in $M''$ ends at $p''$. Choosing any $(q,\ell)$ in
$K'\cap K''$, the corresponding null geodesic will have a
transverse end point in both completions.

We introduce the following topology on $M_\Xi$: since, by
hypothesis, $M$ is paracompact, there exist complete Riemannian
metrics on $M$ ~\cite{NomizuOzeki}. Choose any such metric and let
$d$ be the associated distance function. Similarly $TM$ is
paracompact, and let $\hat d:TM\times TM\to\R^+$ be the distance
function associated to some complete Riemannian metric on $TM$.
Again by paracompactness  there exists a countable dense set
${X_i} \subset NM$, where $NM$ denotes the bundle of null vectors
over $ M$ (see, e.g.~\cite{NomizuOzeki}). Consider the collection
$\cal B$ consisting of the following sets:
\begin{enumerate}
\item All $\hat d$-distance open balls $B_{i,j}\subset M$,  of rational radii $\rho_j$,
centred at the (countable, dense) collection of points $\pi(X_i)$,
where $\pi$ is the projection map from $NM$ to $M$.
(Unsurprisingly, the number $\rho_j$ will be called the radius of
$B_{i,j}$.)
\item All thickenings ${\cal O}_{i,j}$ of null geodesics
generated by those null vectors which belong to $ \hat d$-distance
balls of rational radii $r_j$ around those $X_i$'s which acquire
transverse end points in some extension $\hat \alpha\in \Xi$. Let
$s_i$ be the affine time, in the conformally rescaled metric $\hat
g$, taken by the geodesic generated by $X_i$  to reach the
boundary $\partial \hat M$.  The number $r_{i,j}=\max(s_{i},r_j)$
will be called the radius of ${\cal O}_{i,j}$.
\end{enumerate}

The topology of $M_\Xi$ is defined as the topology generated by
the equivalence classes of sets in $\cal B$ (the reader should
easily check that the intersection condition
of~\cite[p.~78]{Munkres} is satisfied by $\cal B$). Thus, those
equivalence classes provide a countable basis for the topology of
$M_\Xi$. By an abuse of notation we will use the same symbol $\cal
B$ for the collection of equivalence classes.

Now, we have  charts on the set $M_\Xi$, which will give $M_\Xi$
the structure of a manifold with boundary \emph{provided} $M_\Xi$
is a \emph{Hausdorff paracompact} topological space. In order to
establish the Hausdorff property, consider two distinct points
$p_a$, $a=1,2$ in $M_\Xi$, we need to show that they can be
separated by open neighborhoods. The only case that is not
completely straightforward is $p_a\in M_\Xi \setminus M$. Let
$p_1$ arise from a completion $\alpha'$, and let $p_2$ arise from
a completion $\alpha''$ (possibly equal to $\alpha'$), since
$p_1\ne p_2$ we are in case $ii)$ of Definition~\ref{Dugly}, which
implies that sufficiently small coordinate neighborhoods arising
from coordinates near the respective boundaries around the $p_a$'s
will be disjoint.

In order to prove that $M_\Xi$ is a manifold,  it remains to show
that $M_\Xi$ is paracompact. We will show shortly that $M_\Xi$ is
regular as a topological space. Since the topology of $M_\Xi$ has
a countable basis, we can use a theorem of
Urysohn~\cite[Theorem~4.1, p.~217]{Munkres} to conclude that the
topology of $M_\Xi$ is metrisable. Paracompactness follows then
from Smirnov's metrisation theorem~\cite[Theorem~5.1,
p.~260]{Munkres}.

So, we need to establish topological regularity. By
definition~\cite[p.~195]{Munkres}, we need to show that points in
$M_\Xi$ are closed, and that for every $p\in M_\Xi$ and every
closed set $C\subset M_\Xi$ not containing $p$ there exist
disjoint open sets $\cal O$ and $\cal U$ with $p\in \cal O$ and
$C\subset \cal U$. The fact that points are closed is a standard
consequence of the Hausdorff property. On the other hand, the
separation property follows from the existence of local charts, as
follows: let $p \in M_\Xi$, let $C$ be a closed subset of $M_\Xi$,
for each $q\in M_\Xi$ let $U_{i(q)}$ be an element of the basis
$\cal B$ containing $q$, and not containing $p$, with radius
smaller than or equal to $1/i$. Set
$$U_i= \cup_{q\in C}U_{i(q)}\;.$$ Similarly let $V_i$ be an
element of $\cal B$ containing $p$ with radius smaller than or
equal to $1/i$. We wish to show that $V_i\cap U_i=\emptyset$ for
$i$ large enough, and regularity will follow. Suppose that this
last property does not hold, then there exists a sequence of
points $q_i\in C$ such that $U_i(q_i) \cap V_i\ne \emptyset$.
There exists $i_0$ such that for all $i\ge i_0$ all $V_i$'s and
$U_i(q_i)$'s are contained in a single coordinate chart near $p$.
Increasing $i_0$ further if necessary, the $V_i$'s and the
$U_i(q_i)$'s are contained in coordinate balls with radii
approaching zero as $i$ tends to infinity. Furthermore each $V_i$
contains some small coordinate ball centred at $p$. This shows
that for every $j$ there exists $i$ large enough so that
$U_i(q)\subset V_j$ if $U_i(q)\cap V_i\ne\emptyset$, which implies
that $V_j$ intersects $C$, so that $p$ is in the closure of $C$,
hence in $C$, a contradiction.

On $M_\Xi$ we construct a Lorentzian metric as follows: By
paracompactness there exists a covering  $\{{\cal O}_{i}\}_{i \in
\Omega}$  of $M_\Xi$ by local coordinate charts, such that each
chart is a local coordinate chart for some \cbe\
$\alpha_i=(M_i,g_i,\psi_i)\in \Xi$. Again by paracompactness there
exists a partition of unity $\{\varphi_i\}_{i\in\N}$ subordinate
to this covering. On $M$ define
$$
 g_\Xi=\sum_{i\in \N}\varphi_i\psi_i^* g_i\;.
 $$
Note that a convex combination of Lorentzian metrics is not
necessarily a Lorentzian metric; however the $\psi_i^*g_i$'s are
all conformal to each other, which makes it easy to check that
$g_\Xi$ is indeed a Lorentzian metric on $M$, conformal to $g$.

Next, let $M \stackrel{{\hpsi_\Xi}}{\rightarrow}M_\Xi$ be the map
that sends $M$ to the ``$M$" (via the identity) in the disjoint
union that is $M_\Xi$. The metric $g_\Xi$ extends smoothly to the
boundary  for any extension $\hat \alpha$, as a consequence of
Theorem 1, which guarantees that each tensor field
$(\hat\psi^{-1})^*(\varphi_i \psi_i^* g_i)$ smoothly extends from
the interior of $\hat M$  to $\partial \hat M$.
 Using the same symbol $g_\Xi$ for the
metric obtained by extending $g_\Xi$ to the boundary, this makes
$(M_\Xi,g_\Xi, \hpsi_\Xi)$ a \cbe. \qed


\section{Applications}\label{SAppli}
All the results that follow are more or less straightforward
consequences of Theorem~\ref{Tnewmain}. Let us start with the
simplest application (compare~\cite{GerochEspositoWitten}):

\begin{Theorem}
\label{TC1} Suppose that the collection of all future inextendible
null geodesics of $(M,g)$ is simple, then there exists a unique,
up to equivalence, maximal \fcbe\   of $(M,g)$.
\end{Theorem}

\begin{Remark}
The maximal \fcbe\ will be empty if and only if there are no \fcbe
s at all.
\end{Remark}
\proof We let $\Xi$ be the collection of all extensions of
$(M,g)$, and apply Theorem~\ref{Tnewmain}. \qed

\subsection{Strongly causal boundaries, Trautman-Bondi mass}

Clearly, if a null geodesic $\gamma$ terminating at $p\in\partial
\hat M$ is intertwined with some other null geodesic, then strong
causality fails at $p$. Letting $\Xi$ be the collection of
strongly causal \fcbes, it should be clear that the \fcbe\ of
Theorem~\ref{Tnewmain} has a strongly causal boundary, so that we
have:

\begin{Theorem}
\label{Tstrong} Every $(M,g)$ admits a unique, up to equivalence,
 completion which is maximal within the class of
\fcbes\ with strongly causal boundaries.
\end{Theorem}
\begin{Remark}
The maximal \fcbe\ will be empty if and only if there are no \fcbe
s of $(M,g)$ with strongly causal boundaries. Similar comments
apply to our remaining existence results below, and will not be
repeated.
\end{Remark}

As is well known, four-dimensional vacuum space-times with smooth
conformal structure at null infinity can be constructed by solving
a ``hyperboloidal Cauchy problem" (see,
e.g.,~\cite{Friedrich:Pune} and references therein). This leads to
non-empty strongly causal \fcbes, and hence a unique maximal
$\Scri$ by Theorem~\ref{Tstrong}. This remains true in all higher
even space-time dimensions~\cite{AndersonChruscielConformal}.

One of the approaches to the definition of the Trautman-Bondi
mass~\cite{BBM,Tlectures} proceeds via Penrose's conformal
framework~\cite{Geroch:WInicour}. This leads to a potential
ambiguity, related to the possibility of existence of inequivalent
differentiable structures of $\scri$. Our analysis here shows that
for strongly causal conformal completions at infinity no such
ambiguities arise, which establishes well-posedness of the
definitions in~\cite{Geroch:WInicour}. (Compare~\cite{CJL} for an
alternative proof in dimension four.) The conformally-invariant
formula for the Trautman-Bondi mass of cuts of $\Scri$ in higher
dimensions of Ashtekar and Das
\cite{AshtekarDas,HollandsIshibashiMarolf} together with our
result extends to all higher dimensions the four-dimensional
statement here.

%
%

\subsection{Spacelike conformal boundaries}\label{SScb}
Another straightforward application is that to \emph{existence and
uniqueness of maximal spacelike boundaries}:

\begin{Theorem}
\label{Tspace} Every space-time $(M,g)$ admits a unique, up to
equivalence,  \fcbe\ which is maximal within the class of
completions with spacelike boundary.
\end{Theorem}

\begin{Remark}
The result does not, of course, exclude the possibility that there
exist  strictly larger \fcbes\ $\alpha''$; however, any such
$T\partial M''$ will contain non-spacelike tangent spaces.
\end{Remark}

\proof Let $\Omega_\spa$ denote the collection of all null
geodesics which terminate at a point $p$ belonging to some \cbe\
$\hat \alpha$ such that $\partial \hat M$ is spacelike at $p$.
Simplicity of $\Omega_\spa$ is a straightforward consequence of
the next lemma, the result follows then from
Theorem~\ref{Tnewmain} and Remark~\ref{Rnewmain}. \qed

\begin{Lemma}
\label{Lnogood} Suppose that $\gamma_1$ is intertwined with
$\gamma_2$, with $\gamma_1$ terminating at $p\in \partial M'$.
Then $T_p\partial M'$ contains a null direction.
\end{Lemma}

\proof By definition there exists a sequence
$p_i=\psi'(\gamma_2(s_i))$ converging to $p$. Choosing a
subsequence and changing the parameterisation if necessary, the
sequence of null vectors $\psi'_*\dot \gamma_2(s_i)$ converges
then to a null vector tangent to $\partial M'$ at $p$. \qed

\subsection{Null conformal boundaries}\label{SNcb}
An important class of \cbes\ is provided by those for which the
boundary is a null manifold. We note the following corollary of
Lemma~\ref{Lnogood}:

\begin{Proposition}
\label{Pnull} Let $\alpha'$ be a completion of $(M,g)$ with
$\partial M'$ null, and let $\gamma_1$ terminate at $p\in \partial
M'$. Then $\gamma_2$ intertwines $\gamma_1$ if and only if
$\gamma_2$ accumulates at the null generator of $\partial M'$
through $p$.
\end{Proposition}

\proof Let $\gamma$ be the null geodesic generator of $\partial M$
so that $\gamma(0)=p$. We identify $\gamma_a$ with
$\psi'(\gamma_a).$ In order to prove the implication
``$\Rightarrow$", note that the proof of Lemma~\ref{Lnogood} shows
that there exists a parameterisation of $\gamma_2$ so that the
sequence $(\gamma_2(s_i),\dot \gamma_2(s_i))$ approaches
$(\gamma(0),\dot \gamma(0))$ as $i$ tends to infinity, and the
result follows from continuous dependence of geodesics upon
initial values. To prove the reverse implication, consider any
neighborhood $\mcO$ of $p$ such that $(\mcO,g|_\mcO)$ is strongly
causal. Let $s_i$ be an increasing parameter sequence such that
$(\gamma_2(s_i),\dot \gamma_2(s_i))$ approaches $(\gamma(0),\dot
\gamma(0))$ as $i$ tends to infinity. Suppose that $\gamma_2$ is
entirely contained in $\mcO$, standard causality theory shows that
$\gamma_2$ can then be extended by the generator of $\mcO$ at
$\gamma(0)$, hence coincides with this generator near $p$, which
is not possible. It follows that $\gamma_2$ leaves and reenters
all sufficiently small neighborhoods of $p$ an infinite number of
times, and consequently intertwines $\gamma_1$. \qed

\subsection{Simple space-times}\label{Ssst}
Theorem~\ref{TC1} can be sharpened somewhat: Let $\Omega_{\fcb}$
be the collection of all future inextendible null geodesics of
$(M,g)$ satisfying the following property: $\gamma\in
\Omega_{\fcb}$ if and only if there exists some \fcbe\
$\alpha_\gamma$ of $(M,g)$ which terminates $\gamma$. (The
subscript ``fcb" stands for ``future conformal boundary"). This is
clearly an open collection of geodesics. Applying
Theorem~\ref{Tnewmain} to the collection
$$\Xi_\fcb=\{\alpha_\gamma\;,\ {\gamma\in\Omega_{\fcb}}\}$$ of all conformal boundary extensions
we obtain:

\begin{Theorem}
\label{Tnewmain2} There exists a unique, up to equivalence,
 maximal \fcbe\   $(\hat M,\hat g)$ of $(M,g)$ if
and only if $\Omega_{\fcb}$ is simple. In particular, if that last
condition holds, then every \fcbe\   of $(M,g)$ is equivalent to a
subset of $(\hat M, \hat g)$.
\end{Theorem}

We shall say that a manifold $(M,g)$ is \emph{simple to the
future} if $(M,g)$ is conformal to the interior of a manifold
$(M_s,g_s)$ with smooth boundary, with each future inextendible
null geodesic in $(M,g)$ acquiring an end point on the boundary of
$M_s$. This is an obvious extension of a similar definition of
Penrose~\cite{penrose:scri}. Theorem~\ref{Tnewmain2} gives:

\begin{Theorem}
\label{Tugly2} Let $(M,g)$ be simple to the future, then $( M_s,
g_s)$ is the maximal conformal future boundary extension of
$(M,g)$, hence unique up to equivalence.
\end{Theorem}

\subsection{Counting maximal extensions}\label{SCe}
 Let $\mcN_{\mathrm{term}}$ denote the
collection of simple sets $\Omega$ which are maximal in the class
that contains only those geodesics which are terminated by some
\cbe\ of $M$. Theorem~\ref{Tnewmain2} shows that every such
nonempty $\Omega$ defines a unique maximal \cbe. We thus have:
\begin{Theorem}
\label{Tnewmain3} The set of inequivalent maximal \fcbes\   of
$(M,g)$ is in one-to-one correspondence with
$\mcN_{\mathrm{term}}$.
\end{Theorem}

Since any $\Omega(\hat \alpha)$, as defined in
Proposition~\ref{POmst}, can be completed to a maximal simple
$\Omega$,   the number of maximal \fcbes\ of $(M,g)$ is smaller
than or equal to the number of simple maximal collections of
future inextendible geodesics  in $M$. Thus,
Theorem~\ref{Tnewmain3} gives an upper bound on the number of
extensions, without any knowledge about their existence.

\subsection{Cauchy horizons}\label{SCCh}

An important class of \cbes\ is provided by Cauchy horizons. In
that case the boundary is necessarily a null topological
hypersurface. We start with the following:

\begin{Proposition}\label{Pgoodmax}
Consider a globally hyperbolic space-time $(M,g)$, and let $\hat
\alpha$ be a \fcbe\ such that $\partial \hat M$ is null and the
following holds: every future inextendible null geodesic $\gamma$
in $M$ accumulates at $\partial \hat M$. Then $\Omega(\hat
\alpha)$ is maximal.
\end{Proposition}

\proof By hypothesis there exists a sequence $p_i =
\hat\psi(\gamma(s_i))$ such that $p_i\to p$ for some $p\in
\partial \hat M$, reparameterising and passing to a subsequence if
necessary the sequence $\hat \psi_*\dot \gamma(s_i)$ converges to
a null vector $\ell$ at $p$. If $\ell$ is transverse to $\partial
\hat M$, then $p$ is a future end point of $\hat \psi(\gamma)$,
thus $\gamma\in \Omega(\hat \alpha)$. If not, then $\ell$ is
tangent to the generator of $\partial \hat M$ through $p$, and
$\gamma \not \in \Omega(\hat \alpha)$ follows from
Proposition~\ref{Pnull}. \qed

We note:
\begin{Proposition}
\label{PcompactCH} Consider a globally hyperbolic space-time
$(M,g)$, and let $\hat \alpha$ be a \fcbe\ with $\partial \hat M$
null and compact. Then the hypotheses (and hence the conclusions)
of Proposition~\ref{Pgoodmax} hold.
\end{Proposition}

\proof  It is well known that $\hat M$ is diffeomorphic to
\bel{decomp}(-1,1]\times \partial \hat M\;,\ee with every
level set $\{s\}\times \partial \hat M$ being a spacelike Cauchy
surface for $\hat M \setminus \partial \hat M$.  If we denote by
$t$ the projection along the first factor in \eq{decomp}, then
$t\circ \psi$ approaches one along any future inextendible causal
geodesic in $M$, and compactness of $\partial \hat M$ implies that
$\psi(\gamma)$ accumulates at some $p\in
\partial \hat M$. \qed

\subsection{Taub-NUT space-time}\label{subsecTNsts}

\levoca{the wordings needs adapting, refer to where TN is
discussed, perhaps some material from ChI should be included} The
globally hyperbolic region of Taub-NUT space-time $(M,g)$ has two
known inequivalent (but isometric)
\fcbes~\cite{Misner,ChImaxTaubNUT,HajicekTaubNUT}, say $\alpha_1$
and $\alpha_2$, with boundaries $S^3$. We will show that subsets
of $\alpha_1$  can be used to classify all extensions. To fix
notation, we parameterise the Taub region of the Taub-NUT
space-time with $t\in (t_-,t_+)$ and with Euler angles
$(\zeta,\theta,\varphi)$ on $S^3$, with the metric taking the form
\begin{eqnarray} & g= -U^{-1}dt^2 +(2\ell)^2U(d\zeta + \cos\theta
d\varphi)^2 + (t^2 + \ell^2)(d\theta^2+\sin^2\theta d\varphi^2)\,,
\label{UBianchi2}
\end{eqnarray}
where  $$U(t)= \frac {(t_+-t)(t-t_-)}{t^2+\ell^2}\;,$$ $$ t_\pm :=
m\pm \sqrt{m^2+\ell^2}\;.$$ Here $\ell$ and $m$ are strictly
positive constants.

The future extension $\alpha_1$ is obtained as follows: one sets
$M_1= (t_-,t_+]\times S^3$, with metric
\be g_1= -4\ell(d\zeta+\cos\theta d\varphi)dt +(2\ell)^2U(d\zeta + \cos\theta
d\varphi)^2 + (t^2 + \ell^2)(d\theta^2+\sin^2\theta d\varphi^2)\,,
\label{UBianchi2a} \ee with the map $\psi_1$ begin given by the
formula
\bel{psimap1}(t,\zeta,\theta,\varphi)\mapsto (t,
\zeta-(2\ell)^{-1}\int_0^tU^{-1}(x)dx,\theta,\varphi)\;.\ee With
this definition one has $(\psi_1)^*g_1=g$. The extension
$\alpha_2$ is obtained by changing two signs above: thus $M_2=
(t_-,t_+]\times S^3$, with metric
\be  g_2= 4\ell(d\zeta+\cos\theta d\varphi)dt +(2\ell)^2U(d\zeta + \cos\theta
d\varphi)^2 + (t^2 + \ell^2)(d\theta^2+\sin^2\theta d\varphi^2)\,,
\label{UBianchi3} \ee with the map $\psi_2$ being given by
\bel{psimap2}(t,\zeta,\theta,\varphi)\mapsto (t,
\zeta+(2\ell)^{-1}\int_0^tU^{-1}(x)dx,\theta,\varphi)\;,\ee
leading similarly to $(\psi_2)^*g_2=g$. Note that the
transformation $(\zeta,\varphi)\to (-\zeta,-\varphi)$ transforms
$g_1$ into $g_2$, so that  $(M_1,g_1)$ and $(M_2,g_2)$ are
isometric.

The vector field $\partial_\zeta$ is clearly a Killing vector for
all three metrics above, and therefore
$$p_{\parallel}:=g(\partial_\zeta,\dot \gamma)$$ is constant along
the geodesic $\gamma$. Taub space-time has three further Killing
vectors, the exact form of which is irrelevant to us here, they
lead to constants of motion which are denoted by $p_a$, $a=1,2,3$.
It is proved in~\cite{HajicekTaubNUT} (see Theorems 1-3 and the
Appendix there) that:
 \levoca{refer to appendix, or give proofs here}
\begin{enumerate}
\item every inextendible future directed null geodesic with $p_{\parallel}>0$
acquires an end point on $\partial M_1$;
\item similarly every inextendible future directed null geodesic with $p_{\parallel}<0$
is terminated by $\partial M_2$;
\item finally, for every inextendible future directed null geodesic $\gamma(s):[0,s_+)\to M $ with
$p_{\parallel}=0$ the functions $\zeta(s)$, $\theta(s)$ and
$\varphi(s)$ have  finite limits as $s$ tends to the end value
$s_+$.
\end{enumerate}


We will need the following:

\begin{Lemma}
\label{LTN} There exist no extension of the Taub space-time
$(M,g)$ which terminates null geodesics with $p_{\parallel}=0$.
\end{Lemma}

\proof  We will argue by contradiction: let $\gamma:[0,s_+)\to M$
satisfy $p_{\parallel}=0$ and suppose that there exists an
extension $\hat \alpha$ terminating $\gamma$. Then there exists a
neighborhood $\mcO$ of $\dot\gamma (0)$ in $\mcN M$ such that no
null geodesics with initial values in $\mcO$ are intertwined with
$\gamma$. We will show that no such neighborhood exists,
establishing the result.

We thus need to prove that any neighbourhood $\mcO$ of $\dot\gamma
(0)$ contains a null geodesic $\hat \gamma$ which is intertwined
with $\gamma$. We seek to show that an appropriate geodesic  with
$\hat p_\parallel=\epsilon\ne 0$, $\epsilon$ small, will have that
property.

It  follows from the equations in~\cite{MisnerTaub} that the
angular coordinates can always be chosen so that $\sin \theta$ is
strictly bounded away from zero on any given geodesic and on
nearby ones. Recall, further, the relation~\cite{MisnerTaub}
\bel{killrelTN} \sin\theta \cos \varphi p_1+ \sin\theta \sin \varphi
p_2 +\cos \theta p_3 = p_\parallel\;. \ee
 Clearly for any $\epsilon$  we can choose initial data for $\hat \gamma$, near to those for $\gamma$,  so that $\hat
p_i\to p_i$ as $\epsilon\to 0$.

Let $\theta_\infty$ and $\varphi_\infty$ be the limiting values of
$\theta(s)$ and $\varphi(s)$ as $s$ approaches the terminal value
$s_+$ (those exist from what has been said above; in any case this
will be clear from what follows), we want to show that $\hat
\gamma(0)$ can be chosen so that the corresponding values $\hat
\theta_\infty$ and $\hat \varphi_\infty$ for $\hat \gamma$
coincide with those for $\gamma$. The coordinates $\theta$ and
$\varphi$ satisfy the equations~\cite{MisnerTaub}
\begin{eqnarray}
  \frac {d  \varphi (s(t))}{dt} &=& \frac {\hat
p_3}{(t^2+\ell^2)\sin^2\theta\sqrt{
\Big(\frac{p_\parallel}{2\ell}\Big)^2 + U \Big(
 \frac{(p_2\cos \varphi-
p_1\sin\varphi)^2}{t^2+\ell^2}+\frac{p_3^2}{(t^2+\ell^2)\sin^2\theta}\Big)}}
 \;,
 \phantom{xxxxxx} \label{phieq}
 \\
 \frac{d \theta (s(t))}{dt}&=&\frac{p_2\cos \varphi-
p_1\sin\varphi}{(t^2+\ell^2)\sqrt{
\Big(\frac{p_\parallel}{2\ell}\Big)^2 + U \Big(
 \frac{(p_2\cos \varphi-
p_1\sin\varphi)^2}{t^2+\ell^2}+\frac{p_3^2}{(t^2+\ell^2)\sin^2\theta}\Big)}}
  \;, \label{thetadynTN}
\end{eqnarray}
 along $\hat \gamma$,
with the corresponding equations for $\gamma$  obtained by setting
 $ p_\parallel=0$.
%
 We need to make sure that the
value $\hat \varphi_\infty-\varphi_\infty$ approaches zero as
$p_\parallel=\epsilon \to0$; note that the behaviors of the
right-hand-sides in \eq{phieq}-\eq{thetadynTN}  are drastically
different depending upon whether or not $p_\parallel=0$ because
$U$ tends to zero as $t\to t_+$. However, for any $t\ne t_+$ the
right-hand-sides of  \eq{phieq}-\eq{thetadynTN} converge to the
ones with $p_\parallel=0$ as $p_\parallel$ approaches zero.
Further we have the obvious estimate
\bel{Lebest} \left|\frac {d  \varphi (s(t))}{dt}\right| \le \frac
{1}{\sin \theta \sqrt{(t^2+\ell^2)U}}= \frac {1}{\sin \theta
\sqrt{(t_+-t)(t-t_-)}}\;.
 \ee
Similarly
\bel{Lebest2} \left|\frac {d  \theta (s(t))}{dt}\right|\le \frac
{1}{ \sqrt{(t_+-t)(t-t_-)}}\;.
 \ee
  The  functions at the right-hand-sides of  \eq{Lebest}-\eq{Lebest2}
belong to $L^1([t_-,t_+])$, and the conclusion that
$$
 \hat \varphi_\infty-\hat \varphi(0)\to_{\hat p_\parallel \to 0}\varphi_\infty-\varphi(0)\;,\quad
\hat \theta_\infty-\hat \theta(0)\to_{\hat p_\parallel \to
 0}\theta_\infty- \theta(0)
$$
follows from Lebesgue's dominated convergence theorem.

Standard considerations based on the Brouwer fixed point theorem
show that initial values $(\hat \varphi(0),\hat \theta(0))$ can be
found so that
$$
 \hat \varphi_\infty=\varphi_\infty\;,\quad \hat \theta_\infty= \theta_\infty\;,
$$
with  $(\hat \varphi(0),\hat \theta(0))$  being as close to $(
\varphi(0), \theta(0))$  as desired if $\hat p_\parallel$ is
chosen sufficiently small. In conclusion, a sufficiently small
$\hat p_\parallel$ will lead to a $\hat \gamma(0)$ belonging to
$\mcO$, for any neighbourhood $\mcO$ of $\gamma(0)$.

It is convenient to parameterise both $\gamma$ and $\hat \gamma$
by $t$. Since $\hat p_\parallel>0$ the geodesic $\hat \gamma$ has
an end-point on $\partial M_1$, we thus have
\bel{integrapp}(\hat \zeta_1(t), \hat \theta(t),\hat
\varphi(t))\to_{t\to \hat t_+}(\hat
\alpha,\theta_\infty,\varphi_+)\;,\ee for some constant $\hat
\alpha\in \R$, where \bel{zetonedef}\hat \zeta_1(t)=\hat
\zeta(t)-(2\ell)^{-1}\int_0^{t}U^{-1}(x)dx\;.\ee We define
$\zeta_1(t)$ by an obvious equivalent of \eq{zetonedef}. On the
other hand
$$(\zeta(t),
\theta(t),\varphi(t))\to_{t\to t_+}(
\zeta_+,\theta_\infty,\varphi_+)\;.$$ The integral appearing in
\eq{zetonedef} diverges logarithmically as $t\to t_+$, leading to
$$\zeta_1(t)\to_{t\to t_+} -\infty\;.$$Summarising,
$$\theta(t)-\hat \theta(t) \to_{t\to t_+}0\;, \quad\varphi(t)-\hat \varphi(t)\to_{t\to t_+}0\;, \quad \zeta_1(t)-\hat \zeta_1(t)\to_{t\to t_+} -\infty\;.
$$

 Consider, now, any small thickening of $\hat \gamma$ contained in a set
$\mcU$ of the form
$$\{t\in (t_+-\delta,t_+)\;, \zeta_1\in (\hat \alpha-\delta,\hat
\alpha+\delta)\;,\
\theta\in(\theta_\infty-\delta,\theta_\infty+\delta)\;,\
\varphi\in (\varphi_+-\delta,\varphi_++\delta)\}\;,$$ for some
small $\delta$.  Since $\zeta$ and $\zeta_1$ are  periodic
coordinates on $S^3$, both $\zeta$ and $ \zeta_1$ are identified
modulo $2\pi$. This implies that $ \gamma$ will visit and leave
$\mcU$  an infinite number of times, no matter how small $\delta$
is, showing that $\gamma$ and $\hat \gamma$ are intertwined, as
desired.
 \qed

\levoca{one could consider rethinking the organisation} The reader
will note that a simple variation of the  arguments just given
proves the following:

\begin{Proposition}
\label{PTNint} Two future inextendible geodesics $\gamma$ and
$\hat\gamma$ in Taub space-time are intertwined if and only if
\bel{intcond}\lim_{t\to t_+}\varphi(t)-\hat \varphi(t)=\lim_{t\to
t_+}\theta(t)-\hat \theta(t)=0\;, \qquad p_\parallel \hat
p_\parallel \le 0\;,\ p_\parallel \ne \hat p_\parallel\;. \ee
\end{Proposition}

The following provides a classification of boundary extensions of
the Taub region:\footnote{Proposition~\ref{PTaubNUT} also follows
from~\cite[Theorem~5.2]{ChRendall} when the boundary is compact,
because for compact boundaries the action of the connected
component of the group of isometries extends to the boundary. The
proof in that last reference appears to be completely different,
making heavy use of the structure of the isometry group of the
space-time. It should, however, be noted that the large group of
isometries is also used in our proof here, in analysing the
geometry of the set of geodesics.}:

\begin{Proposition}\label{PTaubNUT} Any \fcbe\ $\hat \alpha$ of  Taub space-time, or of a Gowdy space-time,
with a connected boundary is equivalent to a subset of $\alpha_1$
or of $\alpha_2$. In particular $\hat \alpha$ is equivalent to
$\alpha_1$ or $\alpha_2$ when $\partial \hat M$  is compact.
\end{Proposition}

\proof Consider any null geodesic $\gamma$ terminated
transversally by $\partial \hat M$. Lemma~\ref{LTN} shows that
$\gamma$ has a future end point either on $\alpha_1$ or on
$\alpha_2$. Let $\mcU_a\subset
\partial \hat M$ denote the set of end points of null geodesics
which also terminate on $\partial M_a$. Then each $\mcU_a$ is
open, at least one of them is not empty, renaming the $\alpha_a$'s
if necessary it follows that $\mcU_1=\partial \hat M$. This
implies that no geodesics terminated by $\partial \hat M$ and
$\partial M_1$ are intertwined, and we can apply
Theorem~\ref{Tnewmain} with $\Xi=\{\alpha_1,\hat\alpha\}$ to
obtain a new boundary $M_\Xi$ in which both $M_1$ and $\hat M$ are
embedded. Proposition~\ref{PcompactCH} implies that $\alpha_\Xi$
is equivalent to $\alpha_1$, and the result follows. \qed

 We finish this section by providing a large family of examples of
 maximal simple $\Omega$'s. We start with the polarised Gowdy case, a similar
construction applies in Taub space-times. The Gowdy space-times
have $\R\times S^1\times S^1\times  S^1$ topology, with the last
two factors being acted upon by $U(1)\times U(1)$  by isometries
in the obvious way, see e.g.~\cite{ChImaxTaubNUT}. If we denote by
$(t,\theta)$ the coordinates on the first $\R\times S^1$ factor,
and by $(x^a)$ the coordinates on the remaining factors, then
(after a convenient choice of the $x^a$'s) for every
$x=(\theta,x^1)\in S^1\times S^1$ the geodesic $\gamma_L(x)$ lying
in the $(t,x^2)$ plane going to, say, the left, has an end point
in $\partial M_1$ and intertwines  the remaining geodesic in that
plane $\gamma_R(x)$, which has an end-point on $\partial M_2$.
Choose any non-empty open set $\mcU_1\subset \T^2=S^1\times S^1$,
with the $S^1$'s here corresponding to the $\theta$'s and $x^1$'s,
such that the interior of the closure  $\overline {\mcU}_1$ of
$\mcU_1$ coincides with $\mcU_1$, and let $\mcU_2 =\T^2 \setminus
\overline{\mcU}_1$. (Equivalently, $\mcU_1$ and $\mcU_2$ are
disjoint open subsets of $\T^2$ such that $\partial
\mcU_1=\partial \mcU_2$ and $\overline{\mcU}_1\cup \overline
\mcU_2=\T^2$.) Let $\Omega_{\mcU_1}$ be the union of the
collection of sufficiently small neighborhoods (in the space of
inextendible geodesics) of left-going $\gamma_L(x)$ as $x$ runs
over $\mcU_1$ with the collection of small neighborhoods of the
right-going $\gamma_R(x)$ as $x$ runs over $\mcU_2$. It should be
clear that the neighborhoods  can be so chosen that
$\Omega_{\mcU_1}$ is simple, and that any maximal simple
$\hat\Omega_{\mcU_1}$ containing $\Omega_{\mcU_1}$ will satisfy
$\hat \Omega_{\mcU_1}\ne \hat \Omega_{\mcU_1'}$ if $\mcU_1\ne
\mcU_1'$.

An obvious adaptation of the discussion of the last paragraph to
the Taub space-time, together with Proposition~\ref{PTaubNUT},
shows that all maximal future boundary extensions of the Taub
space-time are in one-to-one correspondence with open non-empty
sets $\mcU_1\subset S^2$, where $S^2$ here corresponds to the Hopf
quotient of $\partial M_1\approx S^3$, such that the interior of
the closure  $\overline {\mcU}_1$ of $\mcU_1$ coincides with
$\mcU_1$. If $\pi:S^2\to S^2$ denotes the Hopf projection, then
$\pi^{-1}\mcU_1\subset \partial M_1$ is the set of end-points of
null geodesics terminated by a subset of $\partial M_1$, while
$\pi^{-1}\mcU_2\subset \partial M_2$ is the corresponding subset
of $\partial M_2$.

\section{``Minimally non-Hausdorff" maximal extensions?}
\label{Smqb}

Following~\cite{HajicekNonHausdorff}, a topological space will be
called a $Y$-manifold if  every point has a neighborhood
homeomorphic to an open subset of $\R^n$, with the charts
satisfying the usual smooth compatibility conditions. Thus, one
does not require a $Y$-manifold to be either Hausdorff or
paracompact. An example can be provided by attaching
simultaneously $\partial M_1$ and $\partial M_2$ to the Taub
space, with the obvious charts near the boundaries. It should be
clear from what has been said above that every \cbe\ of Taub-NUT
space with conformal factor one can be viewed as a subset of this
extension.

In general, let $\Xi$ be the collection of all \fcbes\ of $(M,g)$.
Then the $Y$-manifold $M_\Xi$ constructed in the proof of
Theorem~\ref{Tnewmain} provides a $Y$-manifold which resembles
this last non-Hausdorff boundary extension of Taub space. (The
reader will have noticed that the condition of simplicity of
$\cup_{\alpha'\in\Xi}\Omega(\alpha')$ in Theorem~\ref{Tnewmain}
has only been used  to show that $M_\Xi$ is Hausdorff.)

To make this construction useful a better understanding of $M_\Xi$
would be necessary: is $M_\Xi$ paracompact? does it carry a
conformal structure? (note that in Theorem~\ref{Tnewmain} we used
the former to construct the latter). We have not attempted to
analyse those questions. However,  if one is only interested in
Lorentzian extensions, i.e. \cbe s with conformal factor one ---
this is the case when considering e.g.  Cauchy horizons --- then
the existence of a Lorentzian metric on $M_\Xi$ is immediate.

Assuming that $M_\Xi$ can be equipped with a conformal structure,
it is then larger than or equal to any manifold \cbe\ of $(M,g)$
by construction. Moreover, again by construction, it is minimal
with respect to this property. The sense in which this renders
$M_\Xi$ unique needs to be made precise yet.

\def\cprime{$'$} \def\cprime{$'$} \def\cprime{$'$} \def\cprime{$'$}
\providecommand{\bysame}{\leavevmode\hbox
to3em{\hrulefill}\thinspace}
\providecommand{\MR}{\relax\ifhmode\unskip\space\fi MR }
\providecommand{\MRhref}[2]{%
  \href{http://www.ams.org/mathscinet-getitem?mr=#1}{#2}
} \providecommand{\href}[2]{#2}

\end{document}